\documentstyle[12pt]{article}
\if@twoside                                     
\else
\fi
\title{Nonlinear Gauge Transformations and \\
       Exact Solutions of the
       Doebner-Goldin Equation}
\author{P.~Nattermann and
        W.~Scherer\\
  \small Institute for Theoretical Physics A\\[-1ex]
  \small Technical University Clausthal\\[-1ex]
  \small D-38678 Clausthal-Zellerfeld, Germany}
\date{}
\def\Ref#1{(\ref{#1})}
\def\rz{\ifmmode{I\hskip -3pt R}
    \else{\hbox{$I\hskip -3pt R$}}\fi}
\def\cz{\ifmmode{C\hskip-4.8pt\vrule height5.8pt\hskip6.3pt}
    \else{\hbox{$C\hskip-4.8pt\vrule height5.8pt\hskip6.3pt$}}\fi}
\def\gz{\ifmmode{Z\hskip -4.8pt Z}
    \else{\hbox{$Z\hskip -4.8pt Z$}}\fi}                
\def\grad{\vec{\nabla}\!}                       

\let\ds\displaystyle
\def\dt{\partial_t}                             

\def\J{{\vec{J}}}

\def\Im{\mbox{Im}}
\def\Aff#1{\mbox{\it Aff\/}(#1)}

\begin{document}
{\let\newpage\relax
\mbox{}\hfill {\bf ASI-TPA/7/95}\\
\mbox{}\hfill {\sl to appear in ``Nonlinear, Deformed and Irreversible}\\
\mbox{}\hfill {\sl Quantum Systems'', World Scientific (1995)}
\maketitle
}
\begin{abstract}
Invariants of nonlinear gauge transformations of a family of
nonlinear Schr\"o\-din\-ger equations proposed by Doebner and Goldin are used
to characterize the behaviour of exact solutions of these equations.
\end{abstract}

\section{Introduction}
\label{s1}
In this paper we shall exhibit solutions for members of the family of
nonlinear Schr\"odinger equations derived by Doebner and Goldin
\cite{DoeGol1,DoeGol4}. Originally these equations were derived in a
quantum mechanical context and proposed as possible nonlinear quantum
evolution equations.
There are various contributions in this volume dealing with certain
quantum mechanical aspects of this family.
E.g.\ Goldin \cite{Goldin} reviews a derivation of the equation and
discusses gauge transformations of the family, Hennig \cite{Hennig}
gives a geometric
derivation, L\"ucke \cite{Luecke} discusses quantum mechanical
observables in nonlinear evolution equations and, Malomed \cite{Malomed}
et.al.~discuss the stability of plain wave solutions, and Mizrahi and
Dodonov \cite{MizDod} consider an equation generalizing the original
equation in \cite{DoeGol1} and Gaussian solutions thereof.

Owing to the presence of many parameters the above
mentioned family is very large and certain members may be of interest in
other fields of physics as well, e.g.~\cite{Kibble1}--\cite{Bertol1}.
Consequently we shall not confine our interest to only certain parameter
values (e.g. ``small nonlinearity") but attempt to construct solutions
for all values.

The family of nonlinear Schr\"odinger equations \cite{DoeGol4} to
be investigated is parameterized by the
classification parameter $D$ of the unitarily inequivalent group
representations
and five real parameters $D^{\prime}c_1,\ldots,D^{\prime}c_5$:
\begin{equation}
i\hbar\dt\psi
= \left(-\frac{\hbar^2}{2m}\Delta+V(\vec{x})\right)\psi
+ i\frac{\hbar D}{2} \frac{\Delta\rho}{\rho}\psi
+\hbar D^{\prime}
\left(\,\sum_{j=1}^5 c_jR_j[\psi]\right)\psi\,.
\label{dge}
\end{equation}
Here $D^{\prime}$ also has the dimensions of a diffusion
coefficient (so that the $c_j$ are dimensionless), and
the nonlinear functionals $R_j$ are complex homogeneous of
degree zero, defined by:
\begin{displaymath}
\begin{array}{c}
\ds  R_1[\psi] := \frac{\grad\cdot\J}{\rho}\,,\qquad
  R_2[\psi] := \frac{\Delta\rho}{\rho}\,,\qquad
  R_3[\psi] := \frac{\J^{\,2}}{\rho^2}\,, \\[1mm]
\ds  R_4[\psi] := \frac{\J\cdot\grad\rho}{\rho^2}\,,\qquad
  R_5[\psi] := \frac{(\grad\rho)^2}{\rho^2}\,,
\end{array}
\end{displaymath}
where $\rho:=\bar\psi\psi$ and
$\J:=\Im \,(\bar\psi\grad\,\psi) = (m/\hbar)\vec{j}$.
Our analysis is facilitated if we rewrite the family \Ref{dge} wholly in terms
of
densities and currents. Using the expansion of the Laplacian
$\Delta\psi = (iR_1[\psi]+(1/2)R_2[\psi]
- R_3[\psi]-(1/4)R_5[\psi])\psi$, equation \Ref{dge} is of the
general form
\begin{equation}
\label{nse}
{\cal S}_{(\nu, \mu)}(\psi) :=  i\dt\psi - i\sum_{j=1}^2 \nu_j R_j[\psi]\psi
           -\sum_{j=1}^5 \mu_j R_j[\psi]\psi
  - \mu_0 V\psi = 0
\end{equation}
where $\nu =(\nu_1,\nu_2)$ and $\mu = (\mu_0,\dots, \mu_5)$
are real parameters and in particular $\nu_1\neq 0$.

Writing $\psi$ in the form
\begin{equation}
\label{decomp}
  \psi = \exp(\theta_1 + i\theta_2)
\end{equation}
and $\theta = (\theta_1, \theta_2)$ one finds the following equations:
\begin{eqnarray}
\label{ae}
{\cal AP}_{(\nu, \mu)}(\theta)_1 & := & \dt \theta_1 - 2\nu_2 \Delta
\theta_1
- \nu_1 \Delta \theta_2
  - 4 \nu_2 (\grad \theta_1)^2 - 2\nu_1 \grad\theta_1\cdot
  \grad\theta_2 = 0 \\[2mm]
\label{pe}
{\cal AP}_{(\nu, \mu)}(\theta)_2 & := &  \dt \theta_2
  \begin{array}[t]{l}
    \ds +2\mu_2 \Delta \theta_1 + \mu_1 \Delta\theta_2
    + 4(\mu_2+\mu_5)(\grad \theta_1)^2 \\
    \ds + 2(\mu_1+\mu_4)\grad\theta_1\cdot\grad\theta_2
    +\mu_3 (\grad\theta_2)^2 + \mu_0 V = 0\,.
  \end{array}
\end{eqnarray}
We shall refer to \Ref{ae} as the amplitude and to \Ref{pe}
as the phase equation. Due to the ambiguity of the
phase function $\theta_2$ in \Ref{decomp} the nonlinear Schr\"odinger equation
(\ref{nse}) and the amplitude and phase equation
are {\em not} fully
equivalent.
However, any solution of the amplitude and phase equation yields via
(\ref{decomp}) a solution of the nonlinear Schr\"odinger equation
(\ref{nse}),
i.e.
${\cal AP}_{(\nu, \mu)}(\theta) = 0 \Rightarrow
{\cal S}_{(\nu, \mu)}(\psi) = 0$.

\section{Nonlinear gauge transformations}
Before we calculate solutions of the set \Ref{nse} of nonlinear
partial differential
equations (PDEs) \Ref{nse}, let us look for invertible linear
transformations of the real
functions $\theta_1,\,\theta_2$ leaving the structure of the amplitude and
phase equations (\ref{ae},\ref{pe}) invariant, i.e. we are looking for
a matrix $A\in GL(2)$, such that
\begin{equation}
\label{trans}
\theta^\prime(\vec x, t) := \left(
\begin{array}{c}
\ds  \theta_1^\prime(\vec x,t)  \\
\ds  \theta_2^\prime(\vec x,t)
\end{array}
\right)
:= \left(
\begin{array}{c}
\ds A_{11} \theta_1 (\vec x,t)
  +A_{12} \theta_2 (\vec x,t)  \\
  \ds A_{21} \theta_1 (\vec x,t)
  +A_{22} \theta_2 (\vec x,t)
\end{array}
\right)
= A\, \theta(\vec x, t)
\end{equation}
solve some PDE of the set with parameters $(\nu^\prime,
\mu^\prime)$, if
$\theta$ was a solution of (\ref{ae},\ref{pe}).
A short calculation shows that the matrix has to be restricted to
\begin{equation}
\label{aff}
  A (\Lambda, \gamma) = \left( \begin{array}{cc}
      1 & 0 \\
      \gamma & \Lambda \\
      \end{array}\right) \,,
\end{equation}
where $\Lambda\neq 0$, hence to a representation of the one dimensional affine
group \Aff 1. The change of parameters according to this transformation
is given by
\begin{equation}
\label{anumu}
(\nu, \mu) \mapsto (\nu^\prime, \mu^\prime) = A(\Lambda, \gamma).(\nu, \mu)\,,
\end{equation}
where
\begin{equation}
\label{parch}
\begin{array}{c}\ds
  \nu_1^\prime = \frac{\nu_1}{\Lambda}\,,\quad
  \nu_2^\prime = -\frac{\gamma}{2\Lambda}\nu_1 +\nu_2\,,\\[3mm] \ds
  \mu_1^{\,\prime} = -\frac{\gamma}{\Lambda}\nu_1 + \mu_1\,,\quad
  \mu_2^{\,\prime} = \frac{\gamma^2}{2\Lambda}\nu_1-\gamma \nu_2
  - \frac{\gamma}{2}\mu_1+\Lambda \mu_2\,,\quad
  \mu_3^{\,\prime} = \frac{\mu_3}{\Lambda}\,,\\[3mm] \ds
  \mu_4^{\,\prime}= -\frac{\gamma}{\Lambda}\mu_3 + \mu_4\,,\quad
  \mu_5^{\,\prime} = \frac{\gamma^2}{4\Lambda}\mu_3
  - \frac{\gamma}{2}\mu_4
  + \Lambda\mu_5\,,\quad
  \mu_0^{\,\prime} = \Lambda \mu_0\,.
\end{array}
\end{equation}
The corresponding transformation of the wave function
$\psi$ is the one used to linearize a certain subfamily of \Ref{dge} in
\cite{Natter2} and partly in \cite{AubSab1}:
\begin{equation}
\label{nonlintransf}
  N_{A(\Lambda,\gamma)}(\psi) =
  \psi^{\,\frac{1}{2}(1+\Lambda+i\gamma)}
  {\bar\psi}^{\,\frac{1}{2}(1-\Lambda+i\gamma)} \,.
\end{equation}
Since these transformations leave the probability density
$\rho_\psi(\vec x,t) = \psi(\vec x,t)\bar\psi(\vec x,t)$ invariant, they
were called {\em nonlinear gauge transformations} \cite{DoGoNa1} (see also
\cite{Goldin} in this volume).
Starting with a solution of the amplitude and phase equation we can use
transformations $A\in \Aff 1$ to construct solutions of the nonlinear
Schr\"odinger equations of the type (\ref{nse}) by exploiting the
implications represented in the following diagram:
\begin{equation}
\label{di}
\begin{array}{ccc}
{\cal AP}_{(\nu, \mu)}(\theta)  =  0 &
\stackrel{\Ref{trans}}{\Longleftrightarrow} &
{\cal AP}_{A.(\nu, \mu)}(A\,\theta)  =  0 \\[2mm]
\Downarrow & \psi = \exp(\theta_1 + i\theta_2) & \Downarrow
\\[2mm]
{\cal S}_{(\nu, \mu)}(\psi) = 0 & \mbox{} &
{\cal S}_{A.(\nu, \mu)}(N_A(\psi)) = 0
\end{array}
\end{equation}
The change of parameters \Ref{parch} amounts to the action of the
affine group on the eight-parameter space, so it is worthwhile to look for a
parameterization of \Ref{nse} that consists of six parameters invariant
under the action of \Aff 1 and two group parameters. Six functionally
independent invariants have been proposed in \cite{DoGoNa1},
\begin{equation}
\label{gi}
\begin{array}{c}\ds
\iota_1 = \nu_1\mu_2 -\nu_2\mu_1\,,\quad
\iota_2 = \mu_1-2\nu_2\,,\quad
\iota_3 = 1 + \mu_3/\nu_1\,,\quad
\iota_4 = \mu_4-\mu_1\mu_3/\nu_1\,,\\[3mm]\ds
\iota_5 = \nu_1(\mu_2+2\mu_5)-\nu_2(\mu_1+2\mu_4)
    +2\nu_2^2\mu_3/\nu_1\,,\quad
\iota_0 = \nu_1\mu_0\,,
\end{array}
\end{equation}
and as group parameters we might choose $\nu_1\neq 0 $ and $\mu_1$.
Re-expressing the other parameters in terms of the new invariants we get
\begin{equation}
\begin{array}{c}\ds
\nu_2 = \frac{1}{2}(\mu_1-\iota_2)\,,\quad
\mu_2 = \frac{1}{2}\nu_1^{-1}(2\iota_1  -\iota_2\mu_1 +\mu_1^2)\,,\\[3mm] \ds
\mu_3 = (\iota_3-1)\nu_1\,,\quad
\mu_4 =  \iota_4 - \mu_1 + \iota_3 \mu_1\,, \\[3mm]\ds
\mu_5 = \frac{1}{2}\nu_1^{-1}\left(\iota_5 -\iota_1
+ \iota_4(\mu_1-\iota_2) + \frac{1}{2}(\mu_1^2- \iota_2^2)(\iota_3-1)
\right)\,,\quad
\mu_0 = \nu_1^{-1}\iota_0\,.
\end{array}
\end{equation}

For the gauge class containing the linear Schr\"odinger equation \Ref{dge}
($D=D^\prime=0$) the invariants read
\begin{equation}
\label{gilin}
\iota_0 = -\frac{1}{2m}\,,\qquad
\iota_1 = \frac{\hbar^2}{8m^2}\,,\qquad
\iota_j = 0\,\quad j=2,\ldots,5\,,
\end{equation}
such that linear Schr\"odinger  equations are always represented by
invariants of
the type $\iota_0 < 0, \iota_1 >0, \iota_2 =\dots=\iota_5=0$.

For the calculation of solutions of \Ref{nse} we
might limit ourselves to a particular {\em gauge}, i.e.~to a particular
choice of the group parameters $\nu_1,\mu_1$. An appropriate choice is
$\nu_1=1$ and $\mu_1=0$, and it is sufficient to consider the amplitude
and phase equation in this particular ``gauge"
\begin{eqnarray}
\label{aeg}
  \dt \theta_1 & = & -\iota_2 \Delta \theta_1 + \Delta \theta_2
  -2 \iota_2 (\grad \theta_1)^2 + 2 \grad\theta_1\cdot
  \grad\theta_2\,, \\[2mm]
\label{peg}
  \dt \theta_2 & = &
  \begin{array}[t]{l}
    \ds -2\iota_1 \Delta \theta_1
    - \left(2\iota_5+2\iota_1-2\iota_4\iota_2-\iota_2^2(\iota_3-1)\right)
      (\grad \theta_1)^2 \\
    \ds - 2\iota_4 \grad\theta_1\cdot\grad\theta_2
    -(\iota_3-1) (\grad\theta_2)^2 - \iota_0 V \,.
  \end{array}
\end{eqnarray}
We will now exhibit various solutions for these equations along with
the fact that the ``gauge invariant" parameters may be used to
characterize their different behaviour.
\section{Stationary solutions}
Stationarity is defined as usual by $\partial_t\rho = 0$ which is
equivalent to $\partial_t\theta_1 = 0$. Hence, for stationary solutions
the amplitude equation (\ref{aeg}) reduces to
\begin{equation}
\label{aeg_stat}
\Delta ( \iota_2\theta_1 - \theta_2) + \vec\nabla \theta_1\cdot
\vec\nabla (\iota_2\theta_1 - \theta_2) = 0.
\end{equation}
For this equation we have two obvious solutions leading to the following
cases (and subsequent sub-cases).
\subsection{Plane waves}
Plane wave solutions are obtained if $V\equiv 0$ in which case setting
$\theta_1(\vec x, t) = const$ in (\ref{aeg_stat}) yields
$\theta_2(\vec x, t) = \vec k\cdot\vec x + k(t)$ from the amplitude
equation such that the resulting phase equation (\ref{peg}) gives a
dispersion relation:
\begin{equation}
\label{disp}
\partial_t\theta_2 (\vec x, t) = (1-\iota_3)k^2 =
(1+\frac{2mDc_3}{\hbar})k^2 =: \omega(\vec k)
\end{equation}
and the solutions thus found are
\begin{equation}
\label{pw_sol}
\psi(\vec x,t) = \psi_0 \exp\left(i(\vec k \cdot\vec x - \omega(\vec k))\right)
\,,\qquad \psi_0 = const.
\end{equation}
\subsection{Nontrivial stationary solutions}
\label{snt}
Neglecting constant phase factors the Ansatz
\begin{equation}
\label{ansatz2}
\theta_2(\vec x, t) = \iota_2\theta_1(\vec x) - \omega t
\end{equation}
 also solves
(\ref{aeg_stat}) and leads to the following phase equation
\begin{equation}
\label{peg_stat}
2\iota_1\Delta\theta_1 + 2(\iota_1 + \iota_5)(\vec\nabla\theta_1)^2 +
\iota_0 V = \omega\,.
\end{equation}
\subsubsection{Stationary solutions from linear Schr\"odinger equations}
If $\iota_1\neq 0 \neq \iota_1 + \iota_5$ and $\theta_1$ is a solution
of (\ref{peg_stat}) then the function
\begin{equation}
\label{def_phi}
\varphi(\vec x) := \exp\left(
\frac{\iota_1+\iota_5}{\iota_1}\theta_1(\vec x) \right)
\end{equation}
satisfies
\begin{equation}
\label{lin_se}
\frac{2\iota_1^2}{\iota_0(\iota_1+\iota_5)}\Delta \varphi + V\varphi
= \frac{\omega}{\iota_0} \varphi.
\end{equation}
{}From the solutions $\varphi$ of this equation one obtains
the following stationary solutions of the nonlinear equation (\ref{dge})
\begin{equation}
\label{e}
  \psi(\vec x,t) =\left(\varphi(\vec
  x)\right)^\frac{\iota_1}{\iota_1+\iota_5} \exp\left(
  i\frac{\iota_1\iota_2}{\iota_1+\iota_5} \ln [\varphi(\vec x) ]^2
  - i\omega t\right).
\end{equation}
As an example we calculate the ground state of
the harmonic oscillator in one dimension $V(x)= \frac{\kappa}{2} x^2$, for
$\iota_0(\iota_1+\iota_5)<0$
\begin{equation}
\label{nsegr}
  \psi_0 (x,t) = \exp\left(
   -\frac{1}{4}
    \sqrt{\frac{-\kappa\iota_0}{\iota_1+\iota_5} }\, x^2
    +i\left(-\frac{\iota_2}{2}
    \sqrt{\frac{-\kappa\iota_0}{\iota_1+\iota_5} }\, x^2
    -|\iota_1| \sqrt{\frac{-\kappa\iota_0}{\iota_1+\iota_5}}\, t\right)
    \right)\,.
\end{equation}

\subsubsection{Other stationary solutions}
\label{ot_sta_sol}
If $\iota_1 = 0 \neq \iota_5$ the phase equation reduces to
\begin{equation}
\label{pe_stat3}
2\iota_5 \left(\vec\nabla\theta_1\right)^2 + \iota_0 V = \omega\,,
\end{equation}
which obviously can only be solved for potentials bounded from above
($\iota_0>0$) or
below ($\iota_0<0$). For such bounded potential separable in
Cartesian coordinates $V(\vec{x},t) = \sum_{j=1}^n V_j(x_j,t)$ we get
$$
  \theta_1(\vec{x},t) = \sum_{j=1}^n \int^{x_j}
    \sqrt{\frac{C_j-\iota_0 V_j(\xi,t)}{2\iota_5} }d\xi + C \,,\quad
    \sum_{j=1}^n C_j = \omega\,,
$$
or in particular for the free case ($V\equiv 0$)
$\theta_1 (\vec x, t) = \vec k\cdot \vec x + const$ and
$\omega = 2\iota_5 k^2$, hence the solution
\begin{equation}
\label{sol_stat3}
\psi(\vec x, t) = \psi_0 \exp\left(
(1+i\iota_2) \vec k \cdot\vec x - i2\iota_5k^2 t \right)
\,,\qquad \psi_0 = const.
\end{equation}
which is not square integrable.

If on the other hand $\iota_1+\iota_5 = 0 \neq \iota_1$ the phase
equation reduces to
\begin{equation}
\label{pe_stat4}
2\iota_1\Delta\theta_1 = \omega- V\,,
\end{equation}
i.e.\ a Poisson equation for $\theta_1$, which may be solved by a
Poisson integral, if $V(\vec{x},t)$ is suitably decreasing at infinity
$$
  \theta_1(\vec{x},t) = \eta x^2 + \vec{k}\cdot\vec{x} + C
    -(2\iota_1)^{-1} \int_{\rz^n} G_n(\vec x,\vec y) V(\vec{y},t)
    d^n y \,, \qquad \omega= 4\iota_1\eta\,,
$$
where $G_n$ is Green's function for the Poisson equation in $n$
dimensions.
Thus in the free case the solution $\theta_1(\vec x, t) =
\vec{k}\cdot\vec{x} + V+ const$ leads to the wave function
\begin{equation}
\label{sol_stat4}
\psi(\vec x, t) = \psi_0 \exp\left( \eta x^2 + \vec k\cdot\vec x
+2i\iota_2(\eta x^2 + \vec k\cdot\vec x) -4 i\iota_1 \eta
t\right)
\,,\qquad \psi_0 = const
\end{equation}
which is square integrable only if $\eta < 0$.

Finally, if $\iota_1 = 0 = \iota_5$ then the free phase equation is satisfied
by arbitrary $\theta_1$ with $\omega = 0$ and we have the following
stationary solution
\begin{equation}
\label{sol_stat5}
\psi(\vec x, t) = \psi_0 \exp\left( (1 + 2i\iota_2)\theta_1(\vec x)\right)\,,
\qquad \psi_0 = const.
\end{equation}
\section{Gaussian non-stationary solutions}
\subsection{Gaussian wave Ansatz}
In the following we shall consider either the case where a harmonic
potential $V(\vec x) = \frac{\kappa}{2}x^2$ is present or the free
case which is thus described by $\kappa = 0$ \cite{NaScUs1,NaScUs3},
for a special case \cite{DodMiz1}.
Since these cases are separable
we restrict our considerations to one space dimension ($n=1$).
Making a Gaussian wave Ansatz for $\psi$ is equivalent to making the
Ans\"atze
\begin{eqnarray}
\label{gwa1}
\theta_1(x, t) & = & -\frac{(x -s(t))^2}{2\sigma(t)^2}
+ \frac{1}{2}\ln(\sigma(t)) \\
\label{gwa2}
\theta_2(x, t) & = & \iota_0( A(t) x^2 + B(t) x + C(t))\,.
\end{eqnarray}
Inserting these into the amplitude and phase equation (\ref{aeg})
and equating equal powers of $x$ yields coupled nonlinear
ordinary differential equations for $A,B, C, s$ and $\sigma$
as functions of $t$.
This set of  equations can be reduced to two
ordinary second order equations in $\sigma$ and $s$
\begin{eqnarray}
\label{sigtt}
\ddot\sigma & = & \iota_3\frac{(\dot\sigma)^2}{\sigma}
+ 4(\iota_2\iota_3+\iota_4)\frac{\dot\sigma}{\sigma^2}
+8\frac{(\iota_1+\iota_5)}{\sigma^3}
+2\kappa\iota_0\sigma \\
\label{stt}
\ddot s & = & \left( \iota_3\frac{\dot\sigma}{\sigma}
+ 2\frac{(\iota_2\iota_3+\iota_4)}{\sigma^2} \right) \dot s
+ 2\kappa \iota_0 s
\end{eqnarray}
and $A,B$, and $C$ may be expressed in terms of those.
First one has to solve the decoupled equation (\ref{sigtt}) and then
insert the result into (\ref{stt}) and to integrate that equation.
\subsection{General Gaussian solutions}
For the solution of equations \Ref{sigtt} and \Ref{stt} we will treat
the free particle and the harmonic oscillator separately.

In case of the free particle ($\kappa=0$) equation \Ref{sigtt} can be
reduced  by the Ansatz
\begin{equation}
\label{hAn}
  \frac{d}{dt} \sigma^2(t) = h\left(\ln(\sigma^2(t))\right)\,.
\end{equation}
to the following first order differential equation for $h(z)$:
\begin{equation}
\label{hz}
  2 h(z) h^\prime(z) - (1+\iota_3) h(z)^2 -8(\iota_2\iota_3+\iota_4)h(z)
  -32(\iota_1+\iota_5) =0
\end{equation}
Integrating this equation yields in general only an implicit equation
for $h(z)$. In certain special cases we get an explicit expression,
which can be substituted into the Ansatz \Ref{hAn}. As an example let
us treat the case $\iota_1+\iota_5=0$, where
\begin{equation}
\label{hsol1}
  h(z) = \left\{\begin{array}{l@{\qquad}l}
  C_1 \exp\left(\frac{\iota_3+1}{2} z \right)
  -8\frac{\iota_2\iota_3+\iota_4}{1+\iota_3} & \mbox{for } \iota_3\neq
  -1 \,\\
  4(\iota_4-\iota_2)z + C_1           & \mbox{for } \iota_3=-1\,,
  \end{array}\right.
\end{equation}
and we get the implicit solutions
\begin{equation}
\label{sigsol1}
\begin{array}{rcl@{\qquad}l}
  \ds \int_{\sigma_0}^{\sigma(t)} \frac{2xdx}{ C_1 x^{\iota_3+1}
    -8\frac{\iota_2\iota_3+\iota_4}{1+\iota_3}} &=& t-t_0 & \mbox{for
    } \iota_3\neq -1 \,,\\[2mm]
  \ds \int_{\sigma_0}^{\sigma(t)}\frac{2xdx}{8(\iota_4-\iota_2)\ln(x) +  C_1}
&=& t-t_0 &
  \mbox{for } \iota_3 = -1\,.
\end{array}
\end{equation}
 Explicit solutions are obtained for $\iota_3= 1$
\begin{equation}
\label{sigsol2}
  \sigma(t) = \sqrt{ \frac{\exp(2\sigma_1 t-\sigma_1\sigma_2) +
      2(\iota_2+\iota_4)}{\sigma_1}} \,,
\end{equation}
 or for $\iota_4=-\iota_2\iota_3$, $\iota_3\neq 1$
\begin{equation}
\label{sigsol3}
  \sigma(t) = \sigma_0\left(\sigma_1 t +1\right)^{\frac{1}{1-\iota_3}}
\end{equation}
where $\sigma_1, \sigma_2$ are integration constants to be chosen
suitably.

For the harmonic oscillator ($\kappa >0$) we have not found an
analogous Ansatz to reduce \Ref{sigtt}. Instead, for $\iota_3\neq 1$
we transform \Ref{sigtt} to an equation for the newly defined
function
\begin{equation}
\label{defq}
q(t) := \left(  \sigma(t) \right)^{1-\iota_3}
\end{equation}
such that the following two differential equations for $q,s$ describe the
behaviour of the Ansatz (\ref{gwa1}) in (\ref{nse})
\begin{eqnarray}
\label{ddotq}
\ddot q & = & 4(\iota_2\iota_3 + \iota_4) q^{\frac{2}{\iota_3-1}}\dot q -
8(\iota_1+\iota_5)(\iota_3-1) q^{\frac{\iota_3+3}{\iota_3-1}} -
2\kappa\iota_0 (\iota_3-1) q\,, \\
\label{ddots}
\ddot s & = & \left( 2(\iota_2\iota_3 + \iota_4) q^{\frac{2}{\iota_3-1}}
-\frac{\iota_3}{\iota_3-1} \frac{\dot q}{q}\right) \dot s + 2\kappa\iota_0 s
\,.
\end{eqnarray}
The search for solutions of these equations is facilitated if we view
them as one-dimen\-sional Newtonian equations with friction. In this way
we have the potentials
\begin{eqnarray}
\label{qpot}
U_q(q) & = & \left\{
\begin{array}{lcl}
  -32(\iota_1+\iota_5)q
  - 4\kappa\iota_0 q^2
    & \mbox{if} & \iota_3 =  -3  \\[1mm]
  -16(\iota_1+\iota_5)\ln q
  - 2\kappa\iota_0 q^2
    & \mbox{if} & \iota_3 = -1 \\[1mm]
  4\frac{(\iota_1+\iota_5)(\iota_3-1)^2}{\iota_3+1}
     q^{2\frac{\iota_3+1}{\iota_3-1}}
  + \kappa\iota_0(\iota_3-1) q^2
    &\mbox{if}& \iota_3 \neq -3,-1,1
\end{array}
\right. \\
\label{spot}
U_s(s) & = & \kappa\iota_0 s^2
\end{eqnarray}
and the friction forces
\begin{eqnarray}
\label{fric_q}
F_q(q,\dot q) & = & 4(\iota_2\iota_3+\iota_4)q^{\frac{2}{\iota_3-1}}\dot q \\
\label{fric_s}
F_s(\dot s) & = & \left( 2(\iota_2\iota_3 + \iota_4)
q^{\frac{2}{\iota_3-1}}
-\frac{\iota_3}{\iota_3-1} \frac{\dot q}{q}\right) \dot s\,.
\end{eqnarray}
Let us treat the case of the harmonic oscillator ($\kappa>0$) first.
The asymptotic motion of the $q$-particle depends on the potential $U_q$
and the friction force $F_q$. If $\iota_0 < 0 < \iota_1+\iota_5$, the
potential has a minimum. Thus if furthermore $\iota_2\iota_3+\iota_4 < 0$
the friction force $F_q$ breaks the motion, and q tends to the minimum of the
potential and so does $\sigma$:
\begin{equation}
\label{sigasym}
  \lim_{t\to\infty} \sigma(t) = \sigma_\infty :=
    \left(\frac{4(\iota_1+\iota_5)}{-\kappa\iota_0}\right)^{1/4}\,,\qquad
  \lim_{t\to\infty} \dot \sigma(t) = 0 \,.
\end{equation}
As a consequence at some instant the friction force $F_s$ on the $s$ particle
becomes negative and
\begin{equation}
\label{sasym}
  \lim_{t\to\infty} s(t) = 0\,,\qquad \lim_{t\to\infty} \dot s(t) = 0\,.
\end{equation}
The wave function $\psi(t)$ turns out to converge asymptotically to the
ground state of the harmonic oscillator \Ref{nsegr}.
Explicit solutions of \Ref{nse} with this asymptotic behaviour are
quasi-classical states with constant width $\sigma(t) =\sigma_\infty$
(see section \ref{Gsw:sec}).

If the above listed conditions are not fulfilled, then the width of the
Gaussian wave packet might converge to infinity or to zero, i.e. to a
constant or a $\delta$ distribution. Furthermore
this behaviour can depend on the initial condition in the case
$ \iota_1+\iota_5 <0<\iota_0$ when the potential $U_q$ has a maximum
and quasi-classical states with $\sigma(t) = \sigma_\infty$ may again be
taken to illustrate the motion of the wave packet. While the ground state
for $\iota_0 <$ is stable under small perturbations, it is unstable for
$\iota_0 > 0$; a small perturbation leads to a de- or increasing width and
a motion of the centre $s(t)$ away from the origin.

For another discussion of the stability of solutions of \Ref{dge} see
\cite{Malomed} in this volume.

\section{Solitary wave solutions}
\subsection{Gaussian solitary waves}
\label{Gsw:sec}
Gaussian solitary waves are Gaussian waves whose width remains constant:
$\sigma(t)=\sigma_0=const$.
Requiring this to be the case reduces (\ref{sigtt}) to
\begin{equation}
\label{sigtt_sol}
\kappa\iota_0 (\sigma_0)^4 + 4(\iota_1+\iota_5) = 0\,.
\end{equation}
In the free case $(\kappa = 0$) this can only be satisfied if
$\iota_1+\iota_5=0$, for the harmonic oscillator $\sigma_0$ has to be the
width $\sigma_\infty$ \Ref{sigasym} of the ground state. Equation
\Ref{stt} reduces to an ordinary
differential equation of second order:
\begin{equation}
  \ddot s  =  2 (\iota_2\iota_3+\iota_4) \sigma_0^{-2} \dot s
    + 2\kappa \iota_0 s
\end{equation}
For $\kappa > 0$ and $\iota_0<0$ ($\iota_0>0$) this is the equation of motion
of a damped harmonic (anti-)oscillator, for $\kappa=0$ the equation of motion
of a damped free particle. Again the gauge invariant expression
$\iota_2\iota_3+\iota_4$ determines whether we have
damping, pumping or no friction at all.

\subsection{Other solitary waves}
Suppose now $\theta = (\theta_1(\vec x), \theta_2(\vec x,t))$ is a
solution of the free ($V\equiv 0$) equations(\ref{ansatz2}) and
(\ref{peg_stat}) as in section \ref{snt}
and that in addition we have no friction forces for the Gaussian
solutions, $\iota_2\iota_3 +\iota_4=0$, then
it turns out that
\begin{eqnarray}
\label{thetatilde1}
\tilde\theta_1(\vec x,t) & = & \theta_1(\vec x - \vec v t)\,, \\
\label{thetatilde2}
\tilde\theta_2(\vec x,t) & = & \theta_2(\vec x -\vec v t, t)
- \frac{1}{2} \vec v\cdot\vec x +
\frac{v^2}{4}(1-\iota_3) t
\end{eqnarray}
is a {\it non-stationary solution of the amplitude and phase equation
(\ref{aeg}) and (\ref{peg})}. It should be emphasized that the original
equation (\ref{nse}) is only Galilei invariant if $\iota_3=0=\iota_4$
but that the transformations (\ref{thetatilde1}, \ref{thetatilde2}) take
solutions to solutions in the more general case
$\iota_2\iota_3+\iota_4=0$. The resulting solution of (\ref{nse})
is then
\begin{equation}
\label{sol_solit3}
\psi(\vec x,t) = \psi_0\exp\left(
\theta_1(\vec x -\vec vt) + i (\iota_2\theta_1(\vec x -\vec v t)
- \frac{1}{2} \vec v\cdot\vec x +
\frac{v^2}{4}(1-\iota_3) t - \omega t )
\right)
\,,
\end{equation}
where $\theta_1$ is a solution of (\ref{peg_stat}) and $\psi_0=const$.
Consequently the
probability density $\rho(\vec x,t) = \psi(\vec x,t)\bar\psi(\vec
x,t)$ becomes a solitary wave
\begin{equation}
\label{rho_solit}
\rho(\vec x,t) = |\psi_0|^2\exp(2 \theta_1(\vec x -\vec vt))
\end{equation}
moving with constant speed without changing shape.

In particular for $\iota_1(\iota_1+\iota_5)<0$ we get square
integrable solutions of the type
\begin{equation}
\label{sol_cosh}
\begin{array}{rcl}
  \psi(\vec x,t) &=&\ds
 \psi_0 \cosh\left(\vec{k}\cdot(\vec x-\vec v
  t)\right)^{\frac{\iota_1}{\iota_1+\iota_5}} \exp\Bigg\{i\frac{2\iota_2
        \iota_1}{\iota_1+\iota_5} \ln\left[\cosh\left(\vec{k}\cdot (\vec
          x-\vec v t)\right) \right]\\
  &&\ds - 2\frac{\iota_1^2}{\iota_1 +\iota_5} k^2 t\Bigg\}
\end{array}
\end{equation}
\subsection{Solitary waves with arbitrary initial data}
If the parameters are as in the previous section, i.e.
$\iota_2\iota_3+\iota_4=0$ and in addition $\iota_1 = 0 =
\iota_5$ then (as was shown in section \ref{ot_sta_sol}) the phase
equation (\ref{peg_stat}) is satisfied by {\it arbitrary} functions
$\theta_1(\vec x)$. Hence, it follows that for the three parameter
family selected by
the conditions
\begin{equation}
\label{arb_para}
\iota_1 = \iota_2\iota_3 + \iota_4 = \iota_5 = 0
\end{equation}
the nonlinear equation (\ref{nse}) has solutions
\begin{equation}
\label{sol_arb}
\psi(\vec x,t) = \exp\left(
\theta_1(\vec x -\vec vt) + i (\iota_2\theta_1(\vec x -\vec v t)
- \frac{1}{2} \vec v\cdot\vec x +
\frac{v^2}{4}(1+\frac{\iota_4}{\iota_2}) t - \omega t )
\right)
\end{equation}
with {\it arbitrary} $\theta_1(\vec x)$. Just as in the previous section
this leads to solitary wave motion for the ``probability density"
$\rho(\vec x,t)$ but now with arbitrary initial data.
\section{Conclusion}
We have presented numerous explicit solutions for members of the family
of nonlinear equations (\ref{nse}). These solutions are constructed for
the parameters defined by $\nu_1=1, \mu_1=0$ and contain the
parameter combinations which are invariant
under the nonlinear transformation (\ref{nonlintransf}). Using this
transformation and the diagram in (\ref{di}) solutions for all other values of
$\nu_1$
and $\mu_1$ are easily obtained.

The solutions given here contain plane waves, nontrivial stationary
solutions, stationary solutions associated with a linear Schr\"odinger
equation, Gaussian wave packets, and various solitary waves including a
three parameter subfamily of (\ref{nse}) which permits solitary waves
for arbitrary initial data. Recently R.~Zhdanov has found other
explicit solutions of (\ref{nse}) for the case $\iota_3\neq 0 \neq
\iota_5$, $\iota_1=\iota_2=\iota_4=0$ \cite{Zhdano}.

It seems quite remarkable that despite its nonlinear character so many
explicit solutions of equation
(\ref{nse}) can be found. Apart from its possible use in other fields
than quantum mechanics the knowledge of ever more explicit solutions may
help to shed light on the physical meaning (if any) of the various parameters.
\section*{Acknowledgments}
Some of the results presented here were obtained in collaboration with
A. Ushveridze.
We also gratefully acknowledge instructive discussions with
H.-D.~Doebner, G.A.~Goldin, W.~L\"ucke and R.~Zhdanov.

\end{document}